\documentclass[twocolumn,showkeys,showpacs,prb]{revtex4}
\usepackage{graphicx,color}

\begin{document}

\title{Origin of perpendicular magneto-crystalline anisotropy in L1$_0$-FeNi under tetragonal distortion}

\author{Yoshio Miura}
\email{miura@riec.tohoku.ac.jp}

\author{Sho Ozaki}

\author{Yasushi Kuwahara}

\author{Masahito Tsujikawa}

\author{Kazutaka Abe}

\author{Masafumi Shirai}

\affiliation{Research Institute of Electrical Communication (RIEC) and Center for Spintronics Integrated Systems (CSIS), Tohoku University, Katahira 2-1-1, Aoba-ku, Sendai 980-8577, Japan}

\date{\today}

\begin{abstract}
We investigated the origin of perpendicular magneto-crystalline anisotropy (MCA) in L1$_0$ ordered FeNi alloy using first-principles density-functional calculations. We found that the perpendicular MCA of L1$_0$-FeNi arises predominantly from the constituent Fe atoms, which is consistent with recent measurements of the anisotropy of the Fe orbital magnetic moment of L1$_0$-FeNi by x-ray magnetic circular dichroism. Analysis of the second-order perturbation of the spin-orbit interaction indicates that spin-flip excitations between the occupied majority-spin and unoccupied minority-spin bands make a considerable contribution to the perpendicular MCA as does  the spin-conservation term in the minority-spin bands. Furthermore, the MCA energy increases as the in-plane lattice parameter decreases (increasing the axial ratio {\it c/a}). The increase in the MCA energy can be attributed to further enhancement of the spin-flip term due to modulation of the Fe $d(xy)$ and $d(x^2-y^2)$ orbital components around the Fermi level under the  compressive in-plane distortion.
\end{abstract}

\pacs{75.30.Gw, 75.50.Bb, 75.40.Mg}

%\keyword{magneto-crystalline anisotropy, FeNi, a first-principles calculation}

\maketitle

\section{Introduction}

Ferromagnetic materials with strong magneto-crystalline anisotropy (MCA) have attracted much attention for application in high density magnetic recording media and nonvolatile magnetoresistive random access memory (MRAM). In order to ensure sufficient endurance against thermal fluctuations when downsizing a recording bit or memory cell, it is inevitable to adopt  ferromagnetic materials with higher MCA, such as L1$_0$-ordered FePt,\cite{2008Yoshikawa-IEEEtrans} CoPt,\cite{2008_Sakuraba-APL} and {\it D}0$_{22}$-ordered Mn$_{3-\delta}$Ga.\cite{2011Kubota-APL} Furthermore, perpendicularly magnetized electrodes of magnetic tunnel junctions (MTJs) lower the critical current density for current-induced magnetization switching by spin-transfer torque.\cite{2006Meng-APL}

Recently, L1$_0$-ordered FeNi, which is free from rare-earth and/or noble metal elements, have been successfully fabricated as thin films using an alternative monatomic layer deposition technique.\cite{2007Shima-JMMM,2010Mizuguchi-JAP,2011Mizuguchi-JMSJ,2012Kojima-JJAP} The MCA energy of L1$_0$-FeNi thin films depends on the sort of buffer layers  \cite{2011Mizuguchi-JMSJ} as well as the degree of chemical order.\cite{2012Kojima-JJAP} Furthermore, x-ray magnetic circular dichroism (XMCD) and magneto-optical Kerr effect measurements were carried out on alternately layered FeNi thin films grown on Ni/Cu(001) substrates.\cite{2011Sakamaki-APEX} By analyzing the Fe L-edge XMCD spectra using the sum rule,\cite{1992Thole-PRL,1993Carra-PRL} they obtained the anisotropy of the orbital magnetic moment, and found that the Ni-sandwiched Fe layer has a MCA energy of 10 $\mu$eV, while the Fe-sandwiched Ni layer has a positive MCA energy of 60 $\mu$eV.

The MCA of ordered L1$_0$-type alloys {\it TX} ({\it T} = Mn, Fe, Co, Ni, {\it X} = Au, Pt, Pd, Co, Ni) have been theoretically well studied using first-principles density functional calculations.\cite{1995_Mertig-PRB,2000Nakamura-IEEEtrans,2001_Eriksson-RPB,2001Przybylski-JMMM,2004Burkert-PRL,2004Staunton-JPCM,2008Kim-JMMM,2011Mitsumata-JMSJ,2012Kota-JAP} For L1$_0$-type FePt or CoPt, the strong perpendicular MCA was attributed to the Pt atom due to the strong spin-orbit interaction, in which Fe or Co atoms only induce the spin magnetic moment on the Pt site.\cite{1995_Mertig-PRB} On the other hand, L1$_0$-FeNi also shows a perpendicular MCA, despite the weak spin-orbit interaction of Ni compared to that of Pt. Therefore, the physical origin of the perpendicular MCA of L1$_0$-FeNi remains unclear at present. Furthermore, recent XMCD measurements reported the anisotropy of the orbital magnetic moment.\cite{2011Sakamaki-APEX,2012Kotsugi-JMMM} These reports discussed the MCA on the basis of Bruno's relation,\cite{1989Bruno-PRB} in which the MCA energy is proportional to the difference in orbital magnetic moment between two magnetization directions. However, there are additional terms related to spin-flip excitations between the exchange-splitting majority-spin and minority-spin bands in the second-order perturbation of the spin-orbit interaction. The effects of the spin-flip terms on the MCA energies can be neglected for systems with strong exchange coupling, such as a free-standing Fe or Co monolayer,\cite{1993_Freeman-PRB} while the spin-flip term makes a large contribution to MCA for systems including nonmagnetic elements. In fact, previous theoretical work revealed an enhancement of the Pt orbital magnetic moment of L1$_0$-FePt in the in-plane magnetization direction,\cite{2001_Eriksson-RPB,1995_Mertig-PRB} while real-space analysis of the MCA energy indicated a strong perpendicular contribution from Pt atoms.\cite{1995_Mertig-PRB} This means that the spin-flip term is essential to understanding the perpendicular MCA of ordered L1$_0$-type alloys.

In this paper, we discuss the MCA energy of L1$_0$-FeNi theoretically on the basis of first-principles density-functional calculations in order to elucidate the physical origin of perpendicular MCA in L1$_0$-FeNi. In particular, we focus our attention on the spin and atomic-site dependence of the MCA energy using a second-order perturbation of the total energy due to the spin-orbit interaction. Furthermore, we discuss the dependence of the MCA energy on the in-plane lattice constant of L1$_0$-FeNi, providing useful guidelines for the choice of buffer layers in L1$_0$-FeNi thin films to realize larger uniaxial MCA constants $K_{\rm u}$ above 1.0 MJ/m$^3$.

\section{Computational procedure}

\subsection{First-principles calculations}

We carried out first-principles density-functional calculations using the Vienna ab initio simulation package (VASP).\cite{1993Kresse-PRB,1996Kresse-PRB} For the exchange and correlation energy, we adopted the spin-polarized generalized gradient approximation (GGA) proposed by Perdew, Becke, and Ernzerhof (PBE).\cite{1996Perdew-PRL} The nuclei and core electrons were described by the projector augmented wave (PAW) potential \cite{1994PAW-PRB,1999Kresse-PRB} and the wave functions of valence electrons were expanded in a plane-wave basis set with a cutoff energy of 336.9 eV. The $k$-point integration was performed using a modified tetrahedron method with Bl\"ochl corrections \cite{1994Blochl-PRB} with 19800 $k$-points in the first Brillouin zone corresponding to the primitive unit cell of L1$_0$-FeNi. The spin-orbit interaction was included through the force theorem.\cite{1990Daalderop-PRB} We determined the MCA energy from the difference in the sum of energy eigenvalues for magnetizations oriented along the in-plane [100] and out-of-plane [001] directions. The MCA energy was defined to be positive when the magnetic easy axis was perpendicular to the plane. As the unit cell of L1$_0$-FeNi, we adopted the primitive tetragonal structures represented by the solid lines in Fig. 1. The lattice parameter $c$ along the perpendicular ($z$) direction for each in-plane lattice constant $a$  was optimized to minimize the total energy within a tolerance of 10 $\mu$eV. We checked that the MCA energy dependence on the in-plane lattice constants obtained using the VASP-PAW code were well reproduced by all-electron calculations using the full-potential linearized augmented-plane-wave (FLAPW) method encoded in the WIEN2K package.\cite{wien2k}

\begin{figure}[t]
\includegraphics[height=0.17\textheight,width=0.2\textwidth]{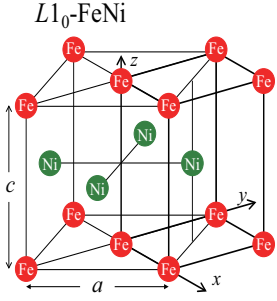}
  \caption{(color online) Schematic figures of the atomic arrangement in L1$_0$-FeNi. The coordinate system and the primitive tetragonal unit cell used in the present study are represented by bold solid lines.}
\end{figure}

\subsection{Second-order perturbation}

In order to obtain a fundamental understanding of the origin of MCA in L1$_0$-FeNi, we consider the second-order perturbation of the total energy due to the spin-orbit interaction $H_{\rm SO}$,\cite{1989Bruno-PRB,1998Laan-JPCM,2006Autes-JPCM} which is given by,
\begin{equation}
E^{(2)} = -\sum_{k} \sum_{n^{\prime}\sigma^{\prime}}^{unocc}  \sum_{n \sigma }^{occ} \frac{|\langle k n^{\prime}\sigma^{\prime}|H_{\rm SO}|kn\sigma\rangle |^2}{\epsilon_{k n^{\prime}\sigma^{\prime}}^{(0)}-\epsilon_{k n \sigma}^{(0)} },
\end{equation}
where $|kn\sigma \rangle$ is an unperturbed state of energy $\epsilon_{k n \sigma}^{(0)}$ with indices of $k$-point $k$, band $n$, and spin $\sigma$. In the tight-binding regime for spin-orbit coupling, $H_{\rm SO}$ is given by the sum of the contributions from each atomic site $i$($j$) due to the localized character of the spin-orbit coupling constant $\xi _i$, 
\begin{equation}
H_{\rm SO}=\sum_{i} \xi _i L \cdot S,
\end{equation}
where $S$ and $L$ are the single-electron spin and angular-momentum operators, respectively. By expanding the $|kn\sigma \rangle$ with an orthogonal basis of atomic orbitals labeled as $\mu$($\lambda$), i.e., $|kn\sigma \rangle=\sum _{i\mu}c_{i\mu \sigma}^{kn}|i\mu \sigma \rangle$,  we can obtain the second order contribution of $H_{\rm SO}$ to the total energy as a sum over terms depending on spin-transition processes, atomic orbitals, and atomic sites,
\begin{eqnarray}
E^{(2)}=-\sum_{\mu \lambda \mu^{\prime} \lambda^{\prime}}\langle \lambda \uparrow|L \cdot S|\lambda^{\prime}\uparrow\rangle \langle \mu^{\prime} \uparrow|L \cdot S|\mu \uparrow\rangle  \nonumber \\ 
\times \sum_{ij} \xi_i \xi_j[G_{\mu \lambda}^{\mu^{\prime} \lambda^{\prime}}(\uparrow,\uparrow;i,j)+G_{\mu \lambda}^{\mu^{\prime} \lambda^{\prime}}(\downarrow,\downarrow;i,j) \nonumber \\
-G_{\mu \lambda}^{\mu^{\prime} \lambda^{\prime}}(\uparrow,\downarrow;i,j)-G_{\mu \lambda}^{\mu^{\prime} \lambda^{\prime}}(\downarrow,\uparrow;i,j)],
\end{eqnarray}
where $G_{\mu \lambda}^{\mu^{\prime} \lambda^{\prime}}(\sigma,\sigma^{\prime};i,j)$ is an integral of joint local density of states (LDOS) given by, 
\begin{eqnarray}
G_{\mu \lambda}^{\mu^{\prime} \lambda^{\prime}}(\sigma,\sigma^{\prime};i,j)=\int_{-\infty}^{E_{\rm F}}d\epsilon \int_{E_{\rm F}}^{\infty}d\epsilon^{\prime} \frac{1}{\epsilon^{\prime} -\epsilon} \nonumber \\
\times \sum_{k}\sum_{n}^{occ}c_{i\mu \sigma}^{kn*} c_{j\lambda \sigma}^{kn} \delta (\epsilon -\epsilon_{kn\sigma}^{(0)}) \nonumber \\
 \sum_{n^{\prime}}^{unocc}c_{i\mu^{\prime}\sigma^{\prime}}^{kn^{\prime}*} c_{j\lambda^{\prime}\sigma^{\prime}}^{kn^{\prime}} \delta(\epsilon^{\prime} -\epsilon_{kn^{\prime}\sigma^{\prime}}^{(0)}),
\end{eqnarray}
where $E_{\rm F}$ is the Fermi energy. The matrix elements of $\langle \mu^{\prime} \uparrow|L \cdot S|\mu \uparrow\rangle $ are given in Ref. \onlinecite{2006Autes-JPCM} as functions of the spherical angular coordinates $\theta$ and $\phi$, which are defined as polar and azimuthal angles between the spin quantization axis and the crystal $z$ axis. The MCA energy is derived as a difference in $E^{(2)}$ between in-plane ($\theta$ = $\pi$/2, $\phi$ = 0) and perpendicular ($\theta$ = 0, $\phi$ = 0) magnetization directions, including the matrix elements $\langle \mu^{\prime} \uparrow|L_{X}|\mu \uparrow\rangle $ and $\langle \mu^{\prime} \uparrow|L_{Z}|\mu \uparrow\rangle $, respectively. We introduce a second-order contribution to the MCA energy depending on the atomic site and the spin-transition process as follows,
\begin{equation}
E_{\rm MCA}^{(2)}=\sum_{i} E_{\rm MCA}^{i} = \sum _{i} \Delta E_{\uparrow \Rightarrow \uparrow}^{i} + E_{\downarrow \Rightarrow \downarrow}^{i} - E_{\uparrow \Rightarrow \downarrow}^{i} - E_{\downarrow \Rightarrow \uparrow}^{i},
\end{equation}
and
\begin{eqnarray}
\Delta E_{\sigma \Rightarrow \sigma^{\prime}}^{i}=-\xi _i \sum_{\mu \lambda \mu^{\prime}\lambda^{\prime}}[\langle \lambda \uparrow |L_{X}|\lambda^{\prime} \uparrow\rangle \langle \mu^{\prime} \uparrow|L_{X}|\mu \uparrow\rangle  \nonumber \\
-\langle \lambda \uparrow|L_{Z}|\lambda^{\prime} \uparrow\rangle \langle \mu^{\prime} \uparrow|L_{Z}|\mu \uparrow\rangle ] \nonumber \\
\times \sum _{j}\xi _j G_{\mu \lambda}^{\mu^{\prime} \lambda^{\prime}}(\sigma,\sigma^{\prime};i,j).
\end{eqnarray}
We directly estimated the second-order perturbative contribution to the spin and atomic site dependent MCA energy $\Delta E_{\sigma \Rightarrow \sigma^{\prime}}^{i}$ by calculating $c_{i\lambda \sigma}^{kn}$ using first-principles calculations without the spin-orbit interaction. $c_{i\lambda \sigma}^{kn}$ can be obtained in the PAW formulation as a sum of the plane-wave part  and the augmented part. Note that $\Delta E_{\sigma \Rightarrow \sigma^{\prime}}^{i}$ in eq. (6) depends quantitatively on the choice of the spin-orbit coupling constant $\xi _i$. We checked that the typical values $\xi _{\rm Fe}$ = 50 meV and $\xi _{\rm Ni}$ = 100 meV \cite{1989Bruno-PRB} in eqs. (3) and (6) were consistent with the MCA energy and orbital magnetic moment of L1$_0$-FeNi obtained from first-principles calculations that included the spin-orbit interaction.

\section{Results}

\begin{figure}[t]
\includegraphics[height=0.25\textheight,width=0.45\textwidth]{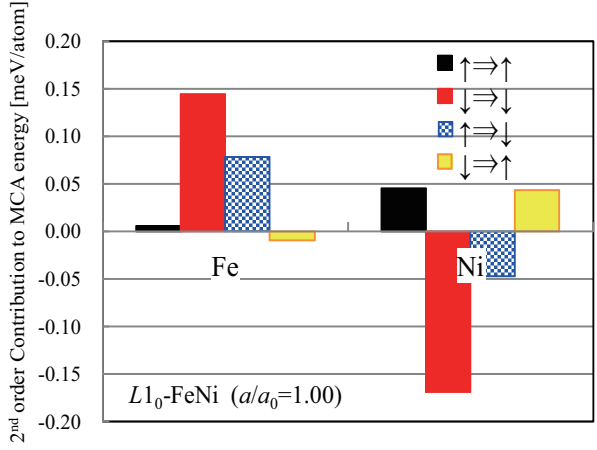}
  \caption{(Color online)  Bar graph of the second-order perturbative contribution to the MCA energy depending on the atomic site and the spin-transition process of L1$_0$-FeNi for an equilibrium lattice constant $a$ = $a_0$ = 3.556 \AA.}
\end{figure}

\subsection{Magneto-crystalline anisotropy and orbital magnetic moment}

We found that the magnetization of L1$_0$-FeNi prefers the [001] direction, with an MCA energy of 0.078 meV per formula unit (f.u.) for the equilibrium lattice constants $a_0$ = 3.556 \AA~and $c_0$ = 3.584 \AA, leading to $K_{\rm u}$ = 0.56 MJ/m$^3$. This value is comparable with experimental observations; i.e., 0.63 MJ/m$^3$ and 0.70 MJ/m$^3.$\cite{2007Shima-JMMM,2011Mizuguchi-JMSJ} Note, however, that the long-range chemical order parameters have been 0.5 or less for samples fabricated so far.\cite{2012Kojima-JJAP} On the other hand, for completely L1$_0$-ordered FeNi, the calculated order parameter value is 1. Experimental observations have indicated that the $K_{\rm u}$ of L1$_0$-FeNi is roughly proportional to the order parameter.\cite{2012Kojima-JJAP} For L1$_0$-ordered FePt, it was observed experimentally that $K_{\rm u}$ increases as a quadratic function of the order parameter.\cite{2002Shima-APL} The same behavior was confirmed in previous theoretical studies based on the coherent-potential approximation (CPA).\cite{2004Staunton-JPCM,2001Przybylski-JMMM,2012Kota-JAP} The spin magnetic moments of the constituent atoms in L1$_0$-FeNi are independent of the magnetization direction, and were evaluated as 2.65 $\mu _{\rm B}$ and 0.61 $\mu _{\rm B}$ for Fe and Ni atoms, respectively. On the other hand, the orbital magnetic moment of each atom shows a characteristic dependence on the magnetization direction. The anisotropy of the orbital magnetic moment defined by $\Delta \mu _{\rm orb}= \mu _{\rm orb}^{[001]} - \mu _{\rm orb}^{[100]}$ takes a positive value, 0.006 $\mu _{\rm B}$, for Fe, whereas it is negative, -0.002 $\mu _{\rm B}$, for Ni in L1$_0$-FeNi.

\begin{figure}[t]
\includegraphics[height=0.4\textheight,width=0.45\textwidth]{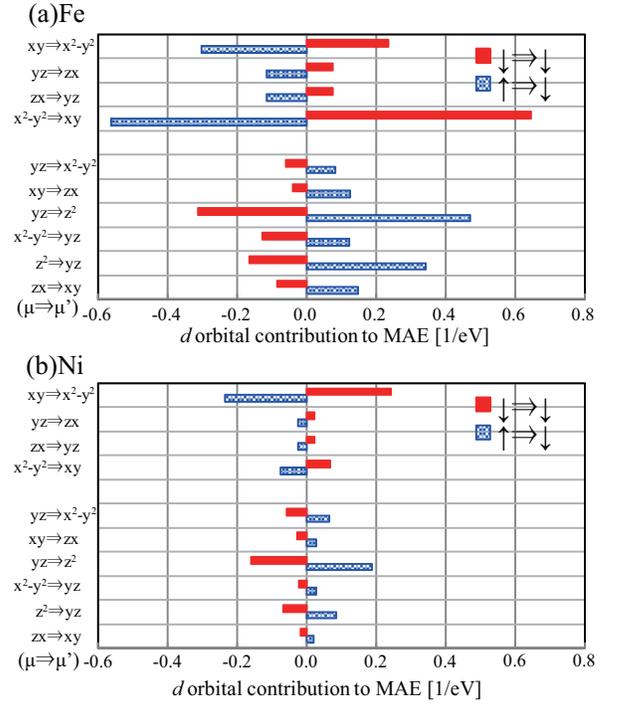}
  \caption{(Color online) Bar graph of the second-order perturbative contribution to the MCA energy of nonvanishing angular momentum matrix elements between $d$ states $\pm |\langle \mu ^{\prime}|L_{X}|\mu\rangle |^2 G_{\mu \mu }^{\mu ^{\prime} \mu ^{\prime}}(\sigma,\downarrow;i,i)$ and $\pm |\langle \mu ^{\prime}|L_{Z}|\mu\rangle |^2 G_{\mu \mu }^{\mu ^{\prime} \mu ^{\prime}}(\sigma,\downarrow;i,i)$ of L1$_0$-FeNi for an equilibrium lattice constant $a$ = $a_0$ = 3.556 \AA.  "$\downarrow \Rightarrow \downarrow$" and "$\uparrow \Rightarrow \downarrow$" indicate spin-conservation and spin-flip terms, respectively.}
\end{figure}

Recently, Kotsugi and his coworkers performed XMCD measurements of alternately layered FeNi thin films grown on Cu$_3$Au buffer layers,\cite{2012Kotsugi-JMMM} They found that the angular distribution in the Fe orbital magnetic moment shows a clear and strong angular dependence, while the Ni orbital magnetic moment displays no clear angular distribution.  This indicates that the anisotropy of the Fe orbital magnetic moment is a main contributing factor to the perpendicular MCA of L1$_0$-FeNi, which is reasonably consistent with the present theoretical results. 

\subsection{Origin of the perpendicular magneto-crystalline anisotropy}

\begin{figure}[t]
\includegraphics[height=0.2\textheight,width=0.45\textwidth]{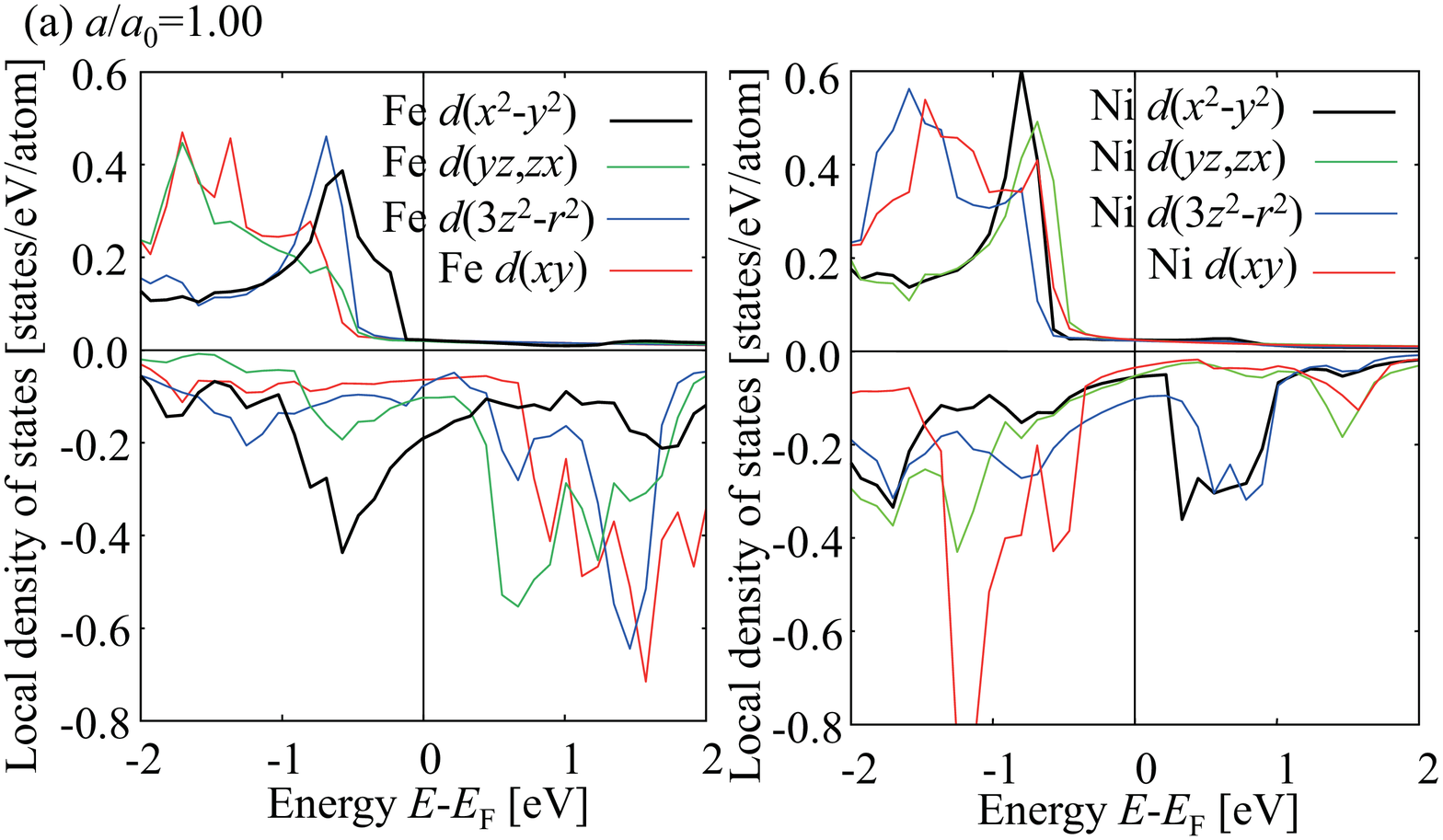}
\includegraphics[height=0.2\textheight,width=0.45\textwidth]{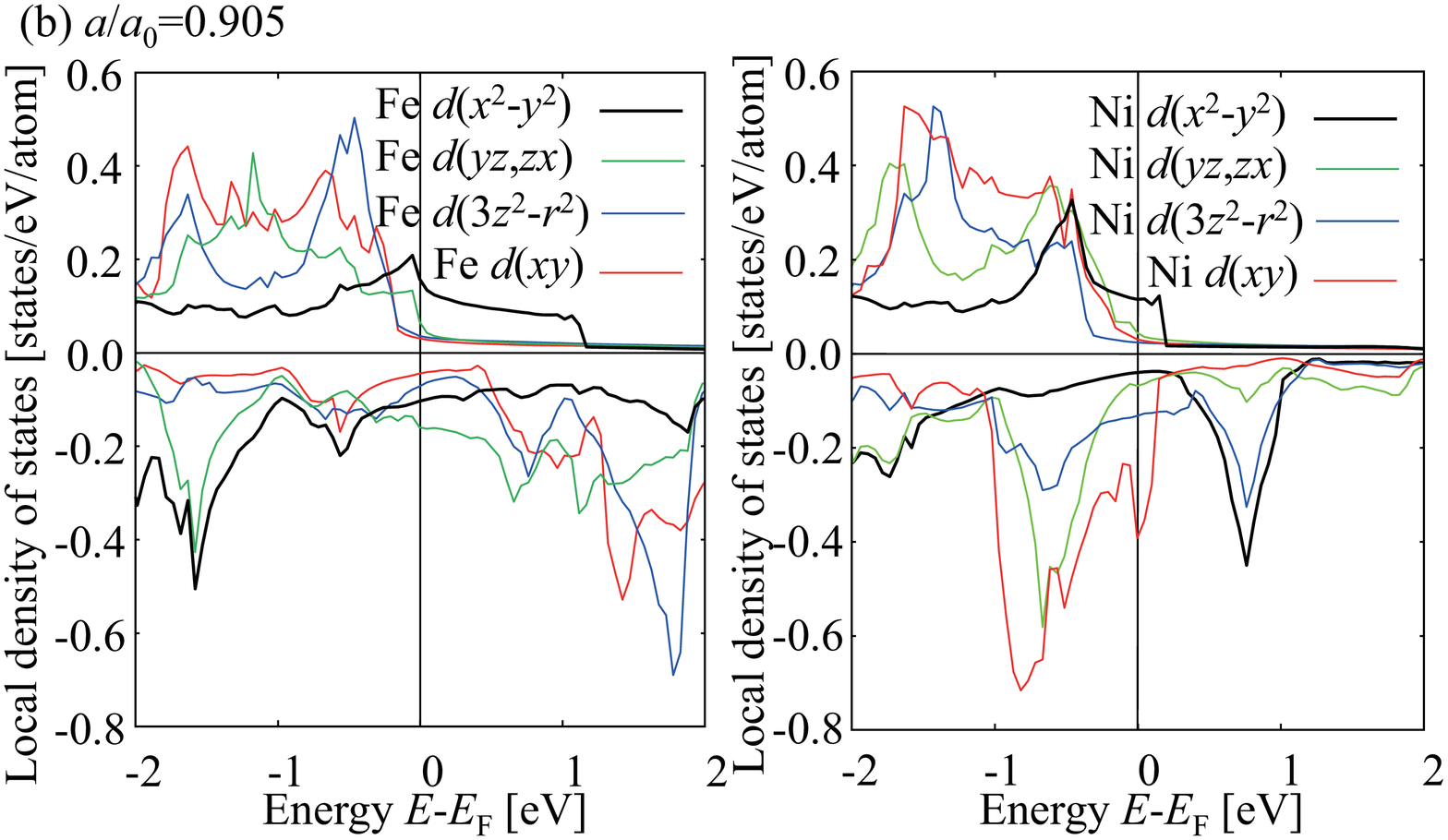}
  \caption{(Color online) Local density of states (LDOS) of the $d$ orbitals of Fe and Ni in L1$_0$- FeNi for (a) $a/a_0$ = 1.00 and (b) $a/a_0$ = 0.904 without spin-orbit interaction as a function of energy relative to the Fermi energy. The equilibrium lattice constant $a_0$ is 3.556 \AA.}
\end{figure}

According to Bruno's relation,\cite{1989Bruno-PRB} the MCA energy is proportional to the anisotropy of the orbital magnetic moment. To confirm a validity of Bruno's relation in L1$_0$-FeNi, we evaluate the second-order perturbative contribution to the MCA energy $\Delta E_{\sigma \Rightarrow \sigma^{\prime}}^{i}$ given in eq. (6). Figure 2 shows a bar graph of $\Delta E_{\sigma \Rightarrow \sigma^{\prime}}^{i}$ for L1$_0$-FeNi in the second-order perturbation. We found that the Fe atom makes a positive contribution to the perpendicular MCA, while the Ni atom makes a negative contribution. These results qualitatively agree with the behavior of the orbital magnetic moment. However, the spin-flip term $\Delta E_{\uparrow \Rightarrow \downarrow}^{\rm Fe}$ also makes a considerable contribution to the perpendicular MCA as does  the spin-conservation term $\Delta E_{\downarrow \Rightarrow \downarrow}^{\rm Fe}$. This means that the evaluation of MCA from the Fe orbital magnetic moment using Bruno's relation\cite{1989Bruno-PRB} will underestimate the MCA energy of L1$_0$-FeNi because of the absence of the spin-flip term in the orbital magnetic moment. The large spin-flip term $\Delta E_{\uparrow \Rightarrow \downarrow}^{\rm Fe}$ can be attributed to the weak exchange splitting of Fe between occupied majority-spin and unoccupied minority-spin states due to hybridization with Ni, resulting in a large majority-spin LDOS around $E_{\rm F}$ (see Fig. 4(a)). The small contributions of $\Delta E_{\uparrow \Rightarrow \uparrow}^{\rm Fe}$ and $\Delta E_{\downarrow \Rightarrow \uparrow}^{\rm Fe}$ to the MCA energy can be attributed to the small LDOS of the unoccupied majority-spin states, which is typical properties of alloys including more-than-half transition metal elements.

\begin{figure}[t]
\includegraphics[height=0.2\textheight,width=0.4\textwidth]{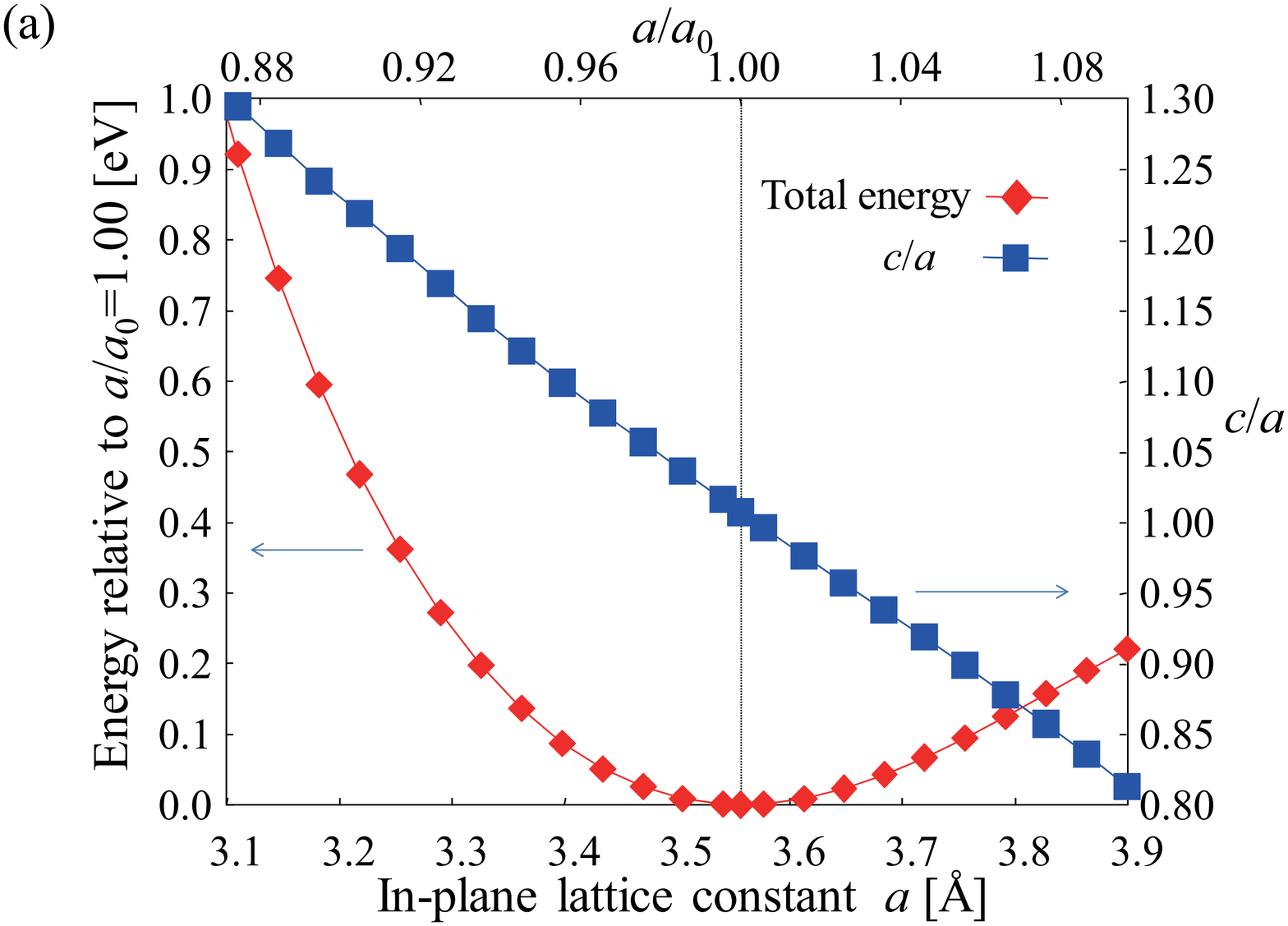}
\includegraphics[height=0.2\textheight,width=0.34\textwidth]{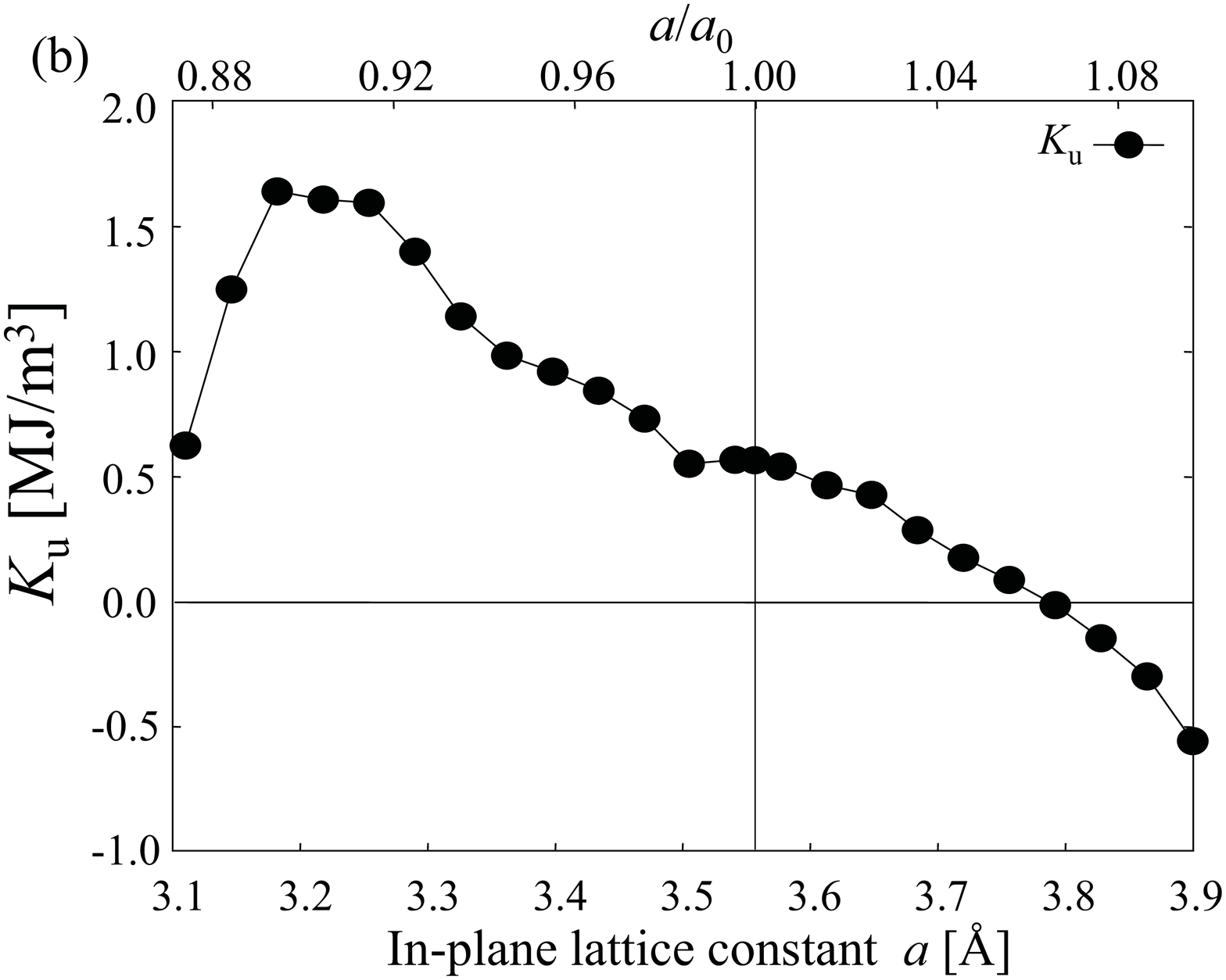}
\includegraphics[height=0.2\textheight,width=0.35\textwidth]{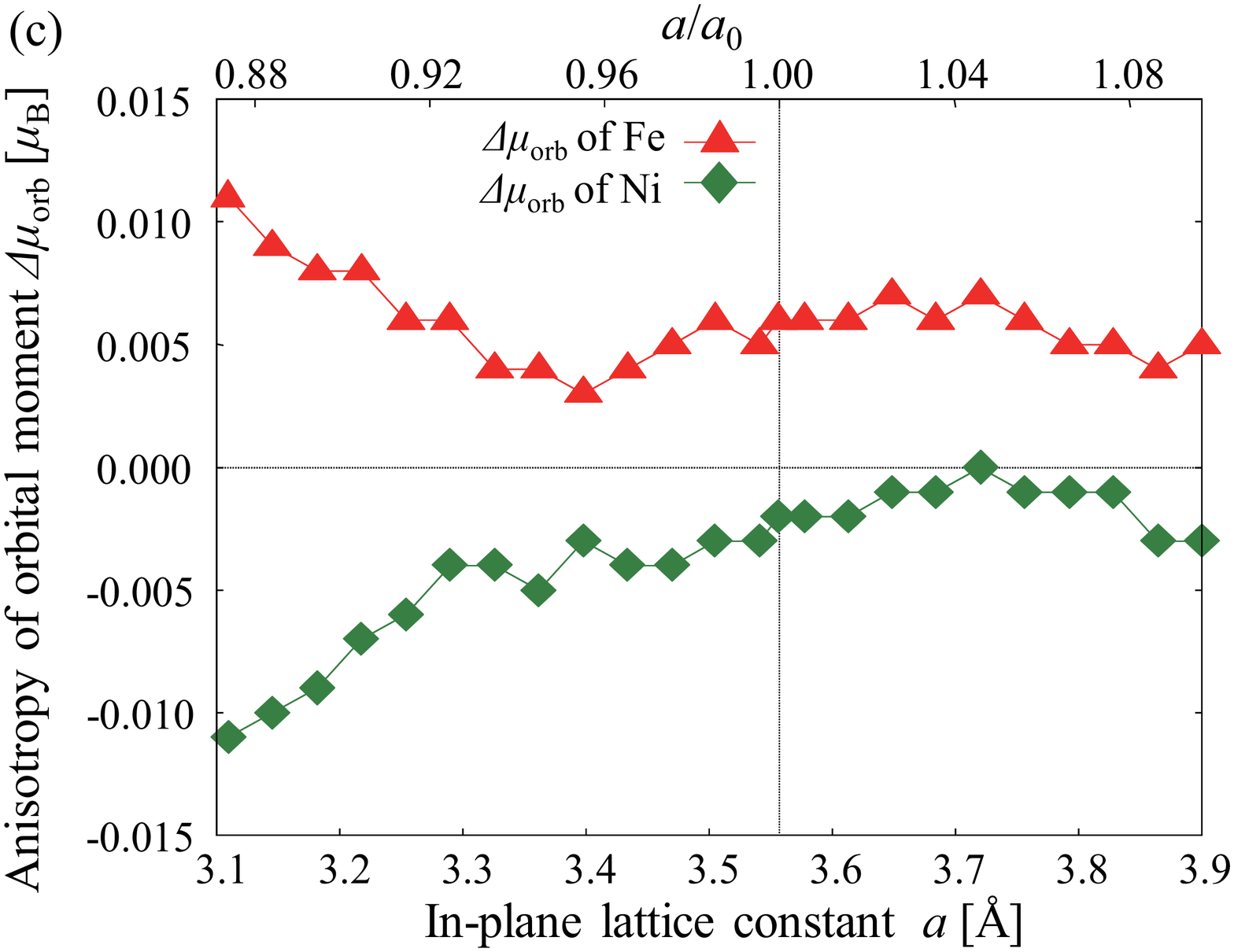}
  \caption{(Color online)  (a) Total energy relative to that of $a/a_0$ = 1.00 (filled red diamonds) and the axial ratio $c/a$ (filled blue squares), (b) the uniaxial MCA constant $K_{\rm u}$ (filled black circles), and (c) the anisotropy of the orbital magnetic moments $\Delta \mu _{\rm orb}= \mu _{\rm orb}^{[001]} - \mu _{\rm orb}^{[100]}$ of constituent Fe and Ni atoms calculated as a function of the in-plane lattice constant $a$ for L1$_0$-FeNi. The equilibrium lattice constant $a_0$ is 3.556 \AA.}
\end{figure}

To obtain a further understanding of the origin of the perpendicular MCA, we show in Fig. 3 the second-order perturbative contribution to the total energy of nonvanishing angular momentum matrix elements between $d$ states, i.e., $\pm |\langle \mu ^{\prime}|L_{X}|\mu\rangle |^2 G_{\mu \mu }^{\mu ^{\prime} \mu ^{\prime}}(\sigma,\downarrow;i,i)$ and $\pm |\langle \mu ^{\prime}|L_{Z}|\mu\rangle |^2 G_{\mu \mu }^{\mu ^{\prime} \mu ^{\prime}}(\sigma,\downarrow;i,i)$. We found large matrix elements  of the spin-conservation term between Fe $d(xy)$ and Fe $d(x^2-y^2)$, which is the main contributing factor to the perpendicular MCA of L1$_0$-FeNi.  Furthermore, the spin-flip term from the majority-spin Fe $d(3z^2-r^2)$ to the minority-spin Fe $d(yz)$ also makes a significant positive contribution to the perpendicular MCA. These results can be confirmed in the Fe LDOS of L1$_0$-FeNi at $a/a_0$ = 1.00 in Fig. 4(a), where there are large LDOS of Fe $d(x^2-y^2)$, $d(3z^2-r^2)$, and $d(yz)$ in the vicinity of $E_{\rm F}$. 

\subsection{Effect of tetragonal distortion}

We examined the dependence of $K_{\rm u}$ on the in-plane lattice constant $a$ of L1$_0$-FeNi, in which the perpendicular lattice parameter $c$ is relaxed for each in-plane lattice constant $a$.  The change in the axial ratio $c/a$ and the total energy are shown in Fig. 5(a). As shown in Fig. 5(b), $K_{\rm u}$ increases with decreasing in-plane lattice constant (with increasing perpendicular lattice parameter $c$), reaching a maximum value of 1.6 MJ/m$^3$ at $a$ = 3.182 \AA. ($c/a$ = 1.242).  The present result suggests that the compressive in-plane stress utilizing the lattice mismatch between L1$_0$-FeNi and the buffer layers is effective in achieving a higher MCA energy. Indeed, previous experiments have indicated that $K_{\rm u}$ monotonically increases with increasing axial ratio $c/a$ of L1$_0$-FeNi.\cite{2011Mizuguchi-JMSJ} According to previous first-principles calculations,\cite{2011Mitsumata-JMSJ} the MCA energy of L1$_0$-FeNi reaches a maximum value of 0.90 MJ/m$^3$ at $c/a$ = 0.95. The discrepancy between this result and that obtained in the present study may partly be attributed to the computational method adopted in the previous study, i.e., the linear muffin-tin orbital method within the atomic sphere approximation, which assumes a spherically symmetric potential in each atomic sphere.

\begin{figure}[t]
\includegraphics[height=0.25\textheight,width=0.38\textwidth]{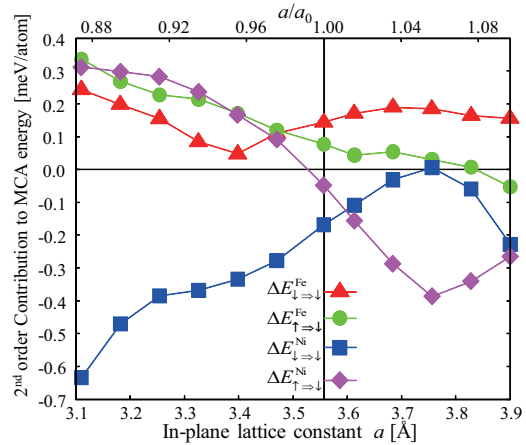}
  \caption{(Color online) The second-order perturbative contribution to the MCA energy $\Delta E_{\downarrow \Rightarrow \downarrow}^{i}$ ($i$=Fe or Ni) as a function of in-plane lattice constant $a$ for L1$_0$-FeNi. The equilibrium lattice constant $a_0$ is 3.556 \AA.}
\end{figure}

\begin{figure}[t]
\includegraphics[height=0.3\textheight,width=0.4\textwidth]{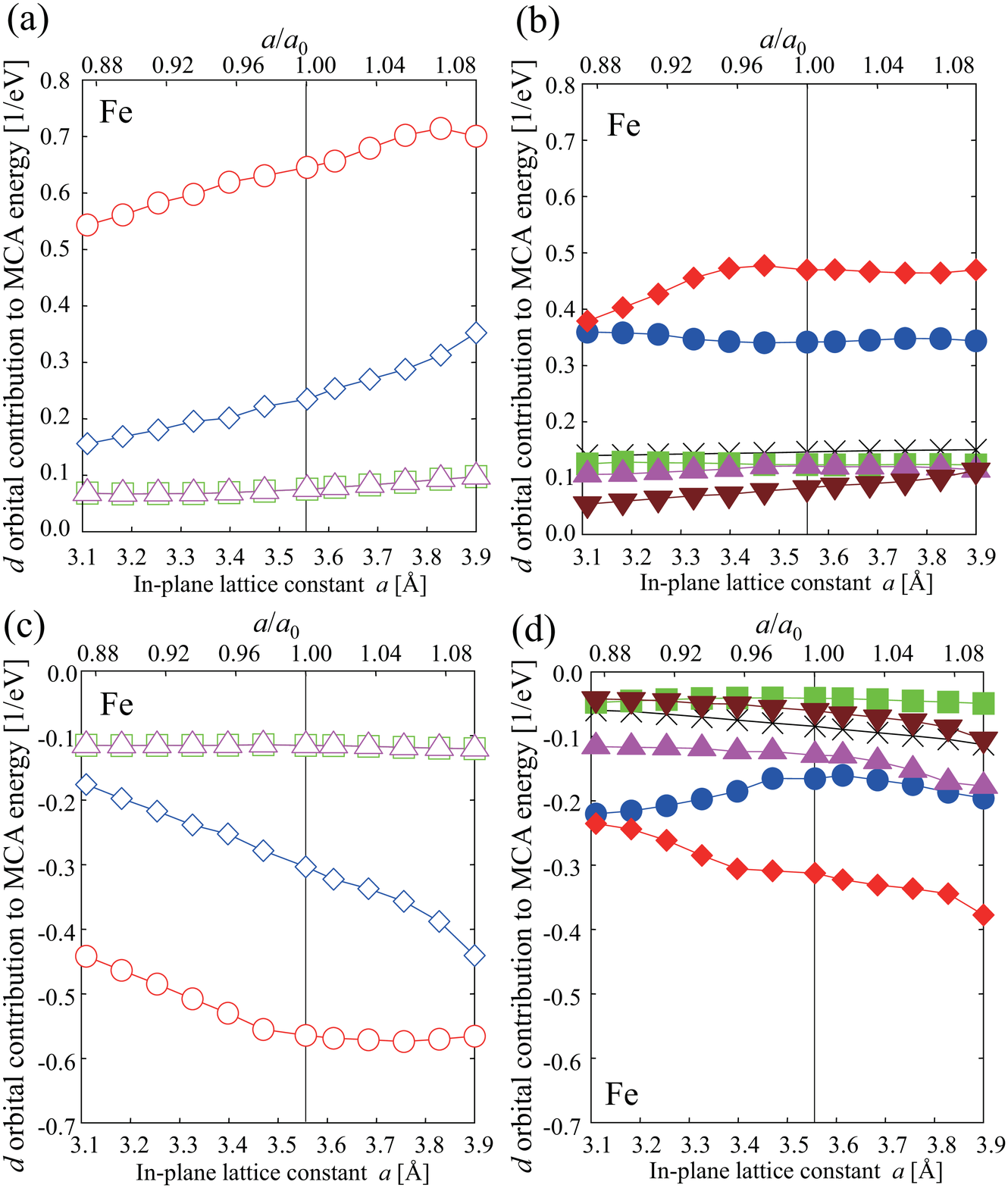}
\includegraphics[height=0.3\textheight,width=0.4\textwidth]{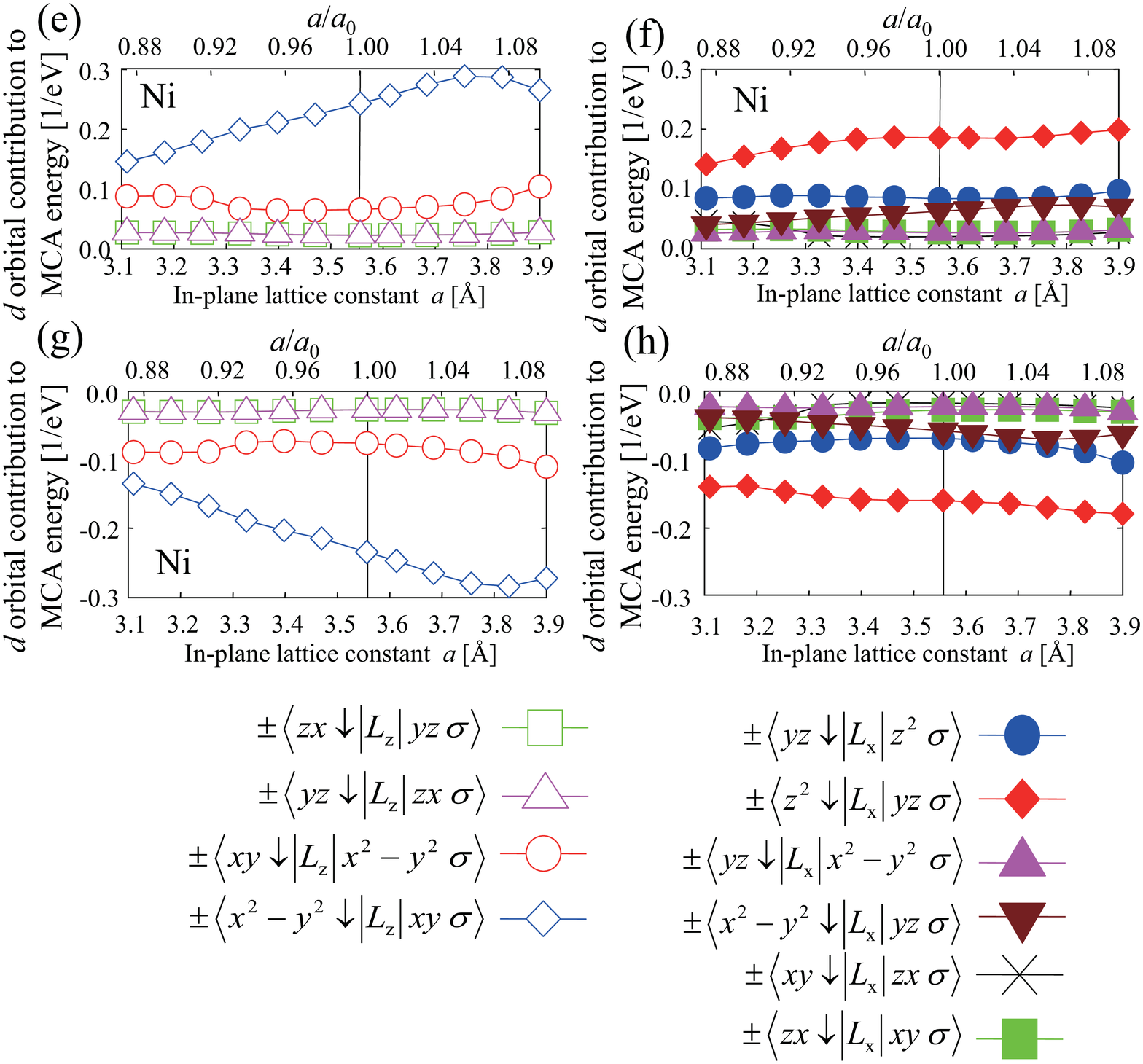}
  \caption{(Color online) The 3$d$ orbital contribution to the MCA energies of L1$_0$-FeNi as a function of the in-plane lattice constant $a$ for Fe (a)$\sim$(d) and Ni (e)$\sim$(g). Here, we use abbreviate notations for the MCA contribution of the nonvanishing angular momentum matrix elements between $d$ states, i.e., $\pm \langle \mu ^{\prime} \downarrow |L_{Z}|\mu \sigma \rangle $ ($"+"$ corresponds to $\sigma=\downarrow$ and $"-"$ to $\sigma=\uparrow$) for $\pm |\langle \mu ^{\prime}|L_{Z}|\mu\rangle |^2 G_{\mu \mu }^{\mu ^{\prime} \mu ^{\prime}}(\sigma,\downarrow;i,i)$, and $\pm \langle \mu ^{\prime} \downarrow |L_{X}|\mu \sigma \rangle $ ($"+"$ corresponds to $\sigma=\uparrow$ and $"-"$ to $\sigma=\downarrow$) for $\pm |\langle \mu ^{\prime}|L_{X}|\mu\rangle |^2 G_{\mu \mu }^{\mu ^{\prime} \mu ^{\prime}}(\sigma,\downarrow;i,i)$. The equilibrium lattice constant $a_0$ is 3.556\AA. }
\end{figure}

To elucidate these results, we show in Fig. 5(c) the anisotropy of the orbital magnetic moments $\Delta \mu _{\rm orb}^{\rm Fe}$ and $\Delta \mu _{\rm orb}^{\rm Ni}$ between cases in which the magnetization is along the [001] or [100] direction as a function of in-plane lattice constant $a$. $\Delta \mu _{\rm orb}^{\rm Fe}$ decreases with decreasing in-plane lattice constant from $a/a_0$ = 1.00, then slightly increases with increasing $a/a_0$. On the other hand, $\Delta \mu _{\rm orb}^{\rm Ni}$ decreases with decreasing in-plane lattice constant from $a/a_0$ = 1.00. These behaviors are inconsistent with the dependence of the MCA energy, indicating that the spin-conservation term is not a primary contributing factor to the large perpendicular MCA of L1$_0$-FeNi under compressive in-plane distortion. To understand this point, we show in Fig. 6 the spin-resolved MCA energy $\Delta E_{\downarrow \Rightarrow \downarrow}^{i}$ and $\Delta E_{\uparrow \Rightarrow \downarrow}^{i}$ as a function of the in-plane lattice constant. First, the spin-conservation terms $\Delta E_{\downarrow \Rightarrow \downarrow}^{\rm Fe}$ and $\Delta E_{\downarrow \Rightarrow \downarrow}^{\rm Ni}$ reproduce the in-plane lattice dependence of the difference in orbital magnetic moments $\Delta \mu _{\rm orb}^{\rm Fe}$ and $\Delta \mu _{\rm orb}^{\rm Ni}$ shown in Fig. 5 (c), because the orbital magnetic moment originates only from the spin-conservation term. Furthermore, the spin-flip terms $\Delta E_{\uparrow \Rightarrow \downarrow}^{\rm Fe}$ and $\Delta E_{\uparrow \Rightarrow \downarrow}^{\rm Ni}$ increase with decreasing in-plane lattice constant, which is consistent with the dependence of the MCA energy shown in Fig. 5(b). This means that the large perpendicular MCA energy of L1$_0$-FeNi under compressive in-plane distortion can be attributed to an increase in the spin-flip terms of Fe and Ni ($\Delta E_{\uparrow \Rightarrow \downarrow}^{\rm Fe}$ and $\Delta E_{\uparrow \Rightarrow \downarrow}^{\rm Ni}$). Figures 7(a)-(h) show the MCA energy of nonvanishing angular momentum matrix elements between $d$ states $\pm |\langle \mu ^{\prime}|L_{X}|\mu\rangle |^2 G_{\mu \mu }^{\mu ^{\prime} \mu ^{\prime}}(\sigma,\downarrow;i,i)$ and $\pm |\langle \mu ^{\prime}|L_{Z}|\mu\rangle |^2 G_{\mu \mu }^{\mu ^{\prime} \mu ^{\prime}}(\sigma,\downarrow;i,i)$ as a function of the in-plane lattice constant. It was found that the positive contribution to the MCA energy decreases, while the negative contribution increases, with increasing in-plane lattice constant. Although these contributions almost cancel each other out, there is a net increase in MCA energy with decreasing in-plane lattice constant. Among them, the matrix elements with $\langle x^2-y^2|L_{Z(X)}|xy\rangle $ and $\langle xy|L_{Z(X)}|x^2-y^2\rangle $ strongly depend on the in-plane lattice constant, in both the spin-conservation and spin-flip terms. We conclude that the significant reduction of the negative contribution related to the matrix elements of $\langle x^2-y^2|L_{Z(X)}|xy\rangle $ and $\langle xy|L_{Z(X)}|x^2-y^2\rangle $ lead to the increase in the MCA energy of L1$_0$-FeNi with decreasing in-plane lattice constant. We confirmed that the LDOS of $d(x^2-y^2)$ and $d(xy)$ are remarkably modulated by compressive in-plane distortion compared with the LDOS of other orbital components. As can be seen in Figs. 4(a) and (b), the LDOS of the majority-spin $d(x^2-y^2)$ below $E_{\rm F}$ are reduced by the tetragonal distortion from $a/a_0$ = 1.00 to $a/a_0$ = 0.905. This can be attributed to an enhancement of the delocalization of the $d(x^2-y^2)$ and $d(xy)$ orbitals resulting from strong bonding between the in-plane $d$ orbitals under the compressive in-plane distortion.

\begin{figure}[b]
\includegraphics[height=0.25\textheight,width=0.38\textwidth]{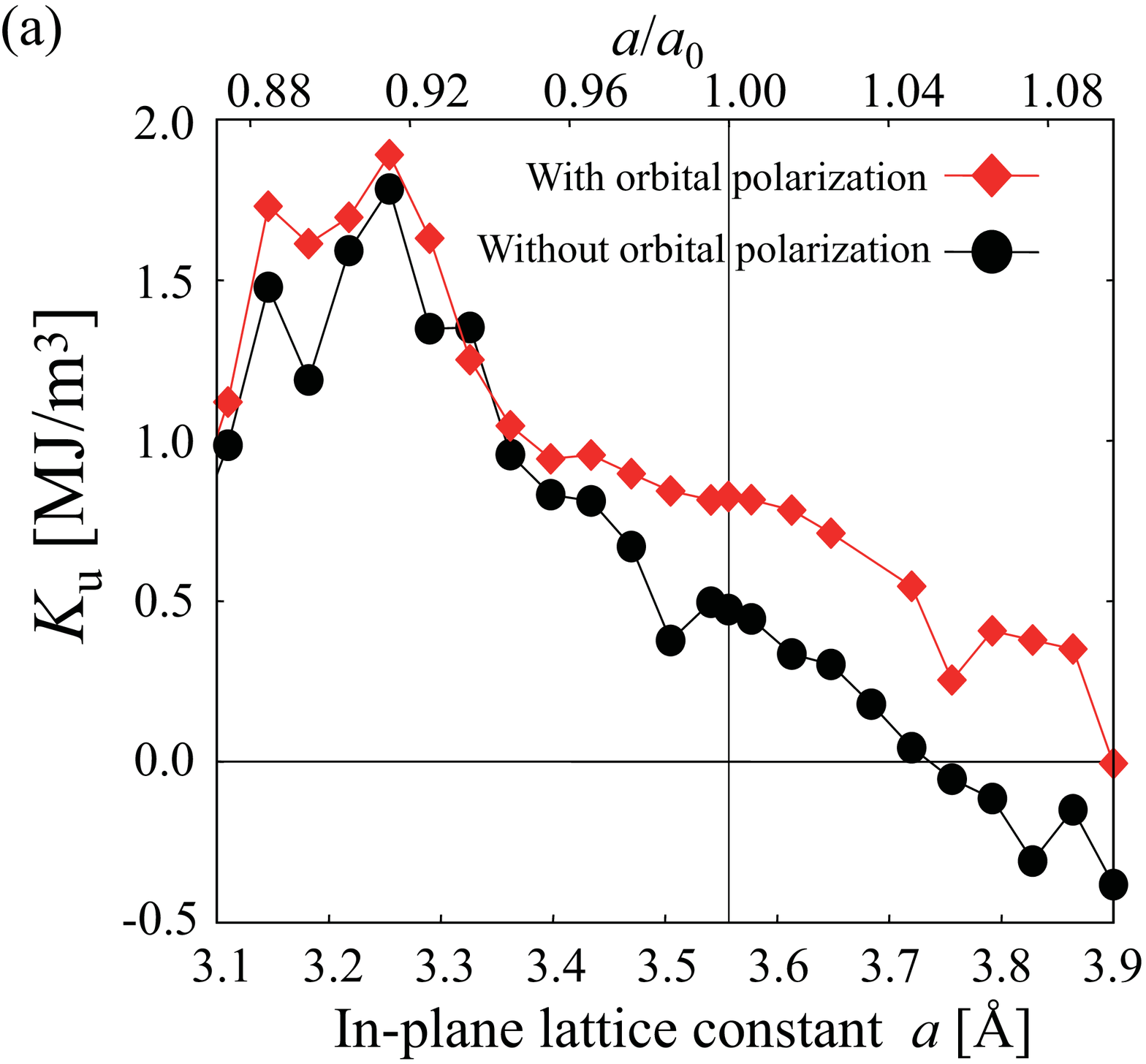}
\includegraphics[height=0.25\textheight,width=0.38\textwidth]{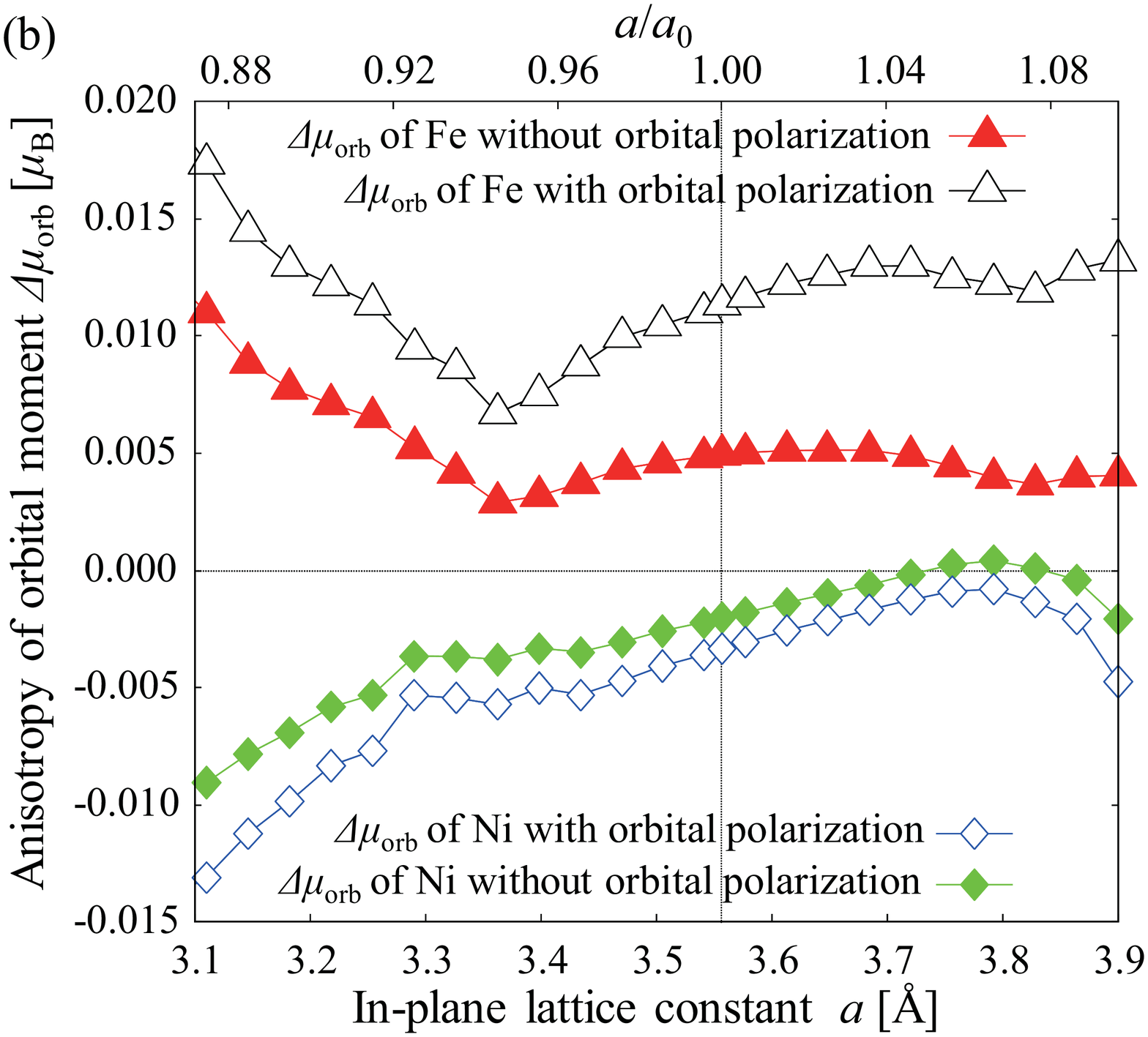}
  \caption{(Color online) (a) The uniaxial magneto-crystalline anisotropy (MCA) energy $K_{\rm u}$ and (b) the anisotropy of the orbital magnetic moments of constituent Fe and Ni atoms $\Delta \mu _{\rm orb}= \mu _{\rm orb}^{[001]} - \mu _{\rm orb}^{[100]}$ with and without the orbital polarization energy as a function of the in-plane lattice parameter $a/a_0$ calculated by the FLAPW method\cite{wien2k}. The equilibrium lattice constant $a_0$ is 3.556 \AA.}
\end{figure}

\section{Discussion}

\subsection{Effect of orbital-polarization}

It is well known that the orbital polarization energy related to the second Hund's law, which is not considered in the present calculation, enhances the MCA energies of most metals and alloys. According to Ravindran {\it et al}.,\cite{2001_Eriksson-RPB} the MCA energy of L1$_0$-FeNi is enhanced from 0.077 meV/f.u. to 0.172 meV/f.u. due to the orbital-polarization energy. To clarify the effect of the orbital polarization (OP) on the MCA energy in the tetragonal distortion of L1$_0$-FeNi, we performed calculations of the MCA energy including the orbital polarization energy using the FLAPW  method with the WIEN2k code\cite{wien2k} using the GGA-PBE function.\cite{1996Perdew-PRL} Figure 8 plots the MCA energy and the anisotropy of the orbital magnetic moment with and without OP as a function of the in-plane lattice constant. At the equilibrium lattice constant ($a/a_0$ = 1.00), we found that the MCA energy with OP is roughly two times that without OP, which is consistent with previous calculation results.\cite{2001_Eriksson-RPB} The origin of the enhancement was well discussed in Ref. \onlinecite{2001_Eriksson-RPB}, i.e., the orbital polarization energy effectively enhances the spin-orbit coupling parameter $\xi _i=\xi _i+BL$, where $B$ is a Racah parameter\cite{1942Racah-PR} expressed in terms of Slater integrals of the single-particle wave functions for the $d$ orbitals. Furthermore, we found that the MCA energy with OP increases with decreasing in-plane lattice constant, which is similar to the results without OP. However, the enhancement of the MCA energy under the compressive in-plane distortion is suppressed compared with the MCA energy without OP. The MCA energies with OP around $a/a_0$ = 0.90 are almost identical to the MCA energies without OP, despite the increase in the anisotropy of the orbital magnetic moment caused by the orbital polarization energy, as shown in Fig. 8(b). This can be better understood by considering the different origin of the perpendicular MCA of L1$_0$-FeNi under tetragonal distortion. As was discussed in the previous section and shown in Fig. 6, the perpendicular MCA around the equilibrium lattice constant ($a/a_0$ = 1.00) is mainly caused by the spin-conservation term of Fe ($\Delta E_{\downarrow \Rightarrow \downarrow}^{\rm Fe}$), while the spin-flip terms of Fe ($\Delta E_{\uparrow \Rightarrow \downarrow}^{\rm Fe}$) and Ni ($\Delta E_{\uparrow \Rightarrow \downarrow}^{\rm Ni}$) provide the largest contribution under highly compressive in-plane distortion ($a/a_0<$0.95). The orbital polarization energy increases the orbital magnetic moment and its anisotropy, which is equivalent to an enhancement of the spin-conservation term, but it does not enhance the spin-flip term because of the absence of a spin operator. Therefore, the effect of the orbital polarization energy is remarkable in the MCA energy arising from the spin-conservation term compared to that from the spin-flip term. 

\subsection{Transport properties through the MgO barrier}

\begin{figure}[b]
\includegraphics[height=0.25\textheight,width=0.45\textwidth]{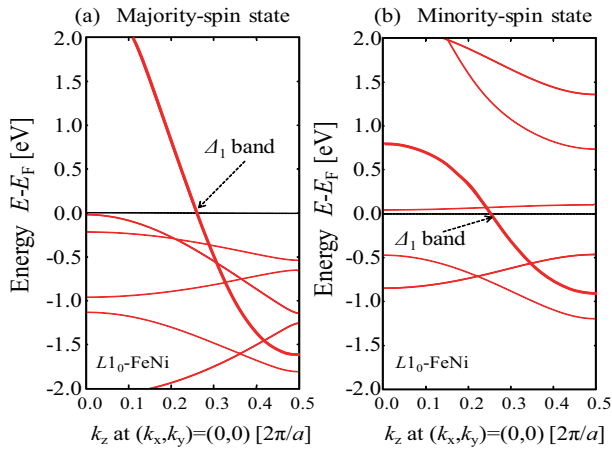}
  \caption{(Color online) Electronic band-dispersion curves relative to the Fermi energy along the [001] direction of bulk L1$_0$-FeNi for (a) majority- and (b) minority-spin states.}
\end{figure}

Finally, we discuss the potential of L1$_0$-FeNi as a ferromagnetic electrode for magnetic tunnel junctions with an MgO barrier. Figure 9 shows the electronic band-structures along the [001] direction of bulk L1$_0$-FeNi. The totally symmetric band in FeNi crosses the Fermi level in both the majority-spin and minority-spin states. The minority-spin $\Delta _1$ band is predominantly composed of Fe and Ni $d(3z^2-r^2)$ orbitals at $E_{\rm F}$, while the majority-spin band is mainly constructed from Fe and Ni $p(z)$ orbitals. Similar results have been obtained for the band dispersion of L1$_0$-FePt.\cite{2008Taniguchi-IEEEtrans} Since the $\Delta _1$ band electrons predominantly transmit  the MgO barrier,\cite{2001Butler-PRB,2001Mathon-PRB} huge TMR ratios cannot be expected in L1$_0$-FeNi/MgO/L1$_0$-FeNi(001) MTJs from the viewpoint of symmetry compatibility between the electronic band structures in bulk FeNi and MgO. In fact, we obtained a very small tunneling magnetoresistance ratio of 51\% for the Fe-terminated L1$_0$-FeNi/MgO/L1$_0$-FeNi(001) MTJ (11\% for the Ni-terminated MTJ) using Landauer-type ballistic conductance calculations. For details regarding these conductance calculations, see ref. \onlinecite{2008Taniguchi-IEEEtrans} and the references therein. Thus, to achieve higher TMR ratios, it is necessary to insert appropriate ferromagnetic layers, e.g. Fe or CoFeB, between the L1$_0$-FeNi electrodes and the MgO barrier.\cite{2012Suzuki-JAP}

\section{Summary}
We evaluated the MCA energy of L1$_0$-FeNi using first-principles density-functional calculations. The perpendicular MCA found in L1$_0$-FeNi can be attributed predominantly to the constituent Fe atoms. The MCA energy of L1$_0$-FeNi increases with decreasing in-plane lattice constant. The perpendicular MCA under tetragonal distortion was elucidated as follows.  The perpendicular MCA at the equilibrium lattice constant is mainly caused by the spin-conservation term in the second-order perturbation of the spin-orbit interaction. On the other hand, the perpendicular MCA under highly compressive in-plane distortion can be attributed to the spin-flip terms of Fe and Ni. The orbital polarization energy increases the MCA energy with a small tetragonal distortion due to enhancement of the spin-conservation term, while the effect of the orbital polarization energy is suppressed under highly compressive in-plane strains. We concluded that modulation of the in-plane lattice parameter of an L1$_0$-FeNi thin film by choosing appropriate buffer layers is effective in obtaining large MCA energies corresponding to 1.0 MJ/m$^3$, and is worth further investigation.

\begin{acknowledgements}
The authors are grateful to M. Mizuguchi, M. Kotsugi, T. Kojima, and K. Takanashi for sharing their experimental data prior to publication. Thanks are also due to Y. Kota, C. Mitsumata, and A. Sakuma for valuable discussion on MCA in disordered systems. This work was partly supported by Collaborative Research Based on Industrial Demand "High Performance Magnets: Towards Innovative Development of Next Generation Magnets" from JST, a Grant-in-Aid for Scientific Research (Grant No. 22246087, No. 22360014 and No. 22760003) from JSPS/MEXT, and the FIRST program from JSPS/CSTP. Y. M. and K. A. gratefully acknowledge support from Mayekawa Houonkai Foundation.
\end{acknowledgements}

\end{document}